\documentclass[fleqn,usenatbib]{mnras}

\usepackage{newtxtext,newtxmath}

\usepackage[T1]{fontenc}

\DeclareRobustCommand{\VAN}[3]{#2}
\let\VANthebibliography\thebibliography
\def\thebibliography{\DeclareRobustCommand{\VAN}[3]{##3}\VANthebibliography}

\usepackage{graphicx}	
\usepackage{amsmath}	
\usepackage{bm}

\usepackage[capitalise]{cleveref}
\crefname{section}{Section}{Sections}
\crefname{figure}{Fig.}{Figs.}
\crefname{equation}{equation}{equations}

\newcommand{\plx}{\bar{\omega}}

\newcommand{\evto}{e_{v_t,\mathrm{(obs)}}}

\newcommand{\uvec}[1]{\bm{\mathit{\hat{#1}}}}

\renewcommand{\vec}[1]{\bm{\mathit{#1}}}

\newcommand{\mat}[1]{\bm{\mathrm{#1}}}      
\newcommand{\matsymb}[1]{\bm{\mathrm{#1}}}  

\newcommand{\transpose}{^\mathrm{T}}

\newcommand{\gaia}{\textit{Gaia}}
\newcommand{\hipparcos}{HIPPARCOS}

\newcommand{\software}[1]{\textsc{\lowercase{#1}}}

\defcitealias{massive}{Paper I}  

\title[Ultramassive white dwarfs]{The origin of ultramassive white dwarfs: hints from \gaia\ EDR3}

\author[Fleury, Caiazzo \& Heyl]{
Leesa Fleury\thanks{email: lfleury@phas.ubc.ca}$^1$,
Ilaria Caiazzo\thanks{email: ilariac@caltech.edu; Sherman Fairchild Fellow}$^2$,
Jeremy Heyl\thanks{email: heyl@phas.ubc.ca}$^1$
\\
$^{1}$Department of Physics and Astronomy, University of British Columbia, Vancouver, BC V6T 1Z1, Canada\\
$^{2}$TAPIR, Walter Burke Institute for Theoretical Physics, Mail Code 350-17, Caltech, Pasadena, CA 91125, USA
}

\date{Accepted XXX. Received YYY; in original form ZZZ}

\pubyear{2023}

\begin{document}
\label{firstpage}
\pagerange{\pageref{firstpage}--\pageref{lastpage}}
\maketitle

\begin{abstract}
\gaia\ Data Release 2 revealed a population of ultramassive white dwarfs on the Q branch that are moving anomalously fast for a local disc population with their young photometric ages.
As the velocity dispersion of stars in the local disc increases with age, a proposed explanation of these white dwarfs is that they experience a cooling delay that causes current cooling models to infer photometric ages much younger than their true ages.
To explore this explanation, we investigate the kinematics of ultramassive white dwarfs within 200 pc of the Sun using the improved \gaia\ Early Data Release 3 observations.
We analyse the transverse motions of $0.95 - 1.25~M_\odot$ white dwarfs, subdivided by mass and age, and determine the distributions of the three-dimensional components of the transverse velocities.
The results are compared to expectations based on observed kinematics of local main-sequence stars.
We find a population of photometrically young ($\sim 0.5 - 1.5$~Gyr) ultramassive ($\sim 1.15 - 1.25~M_\odot$) white dwarfs for which the transverse velocity component in the direction of Galactic rotation is more dispersed than for local disc stars of any age; thus, it is too dispersed to be explained by any cooling delay in white dwarfs originating from the local disc.
Furthermore, the dispersion ratio of the velocity components in the Galactic plane for this population is also inconsistent with a local disc origin. 
We discuss some possible explanations of this kinematically anomalous population, such as a halo origin or production through dynamical effects of stellar triple systems.
\end{abstract}




\section{Introduction}

The kinematics of stars can encode important information about their history and origin.
The dynamical history of the Galaxy is imprinted on the kinematics-age relations of stars in the solar neighbourhood, such as the trend for the velocity dispersion of stars from the local Galactic disc to increase with age due to disc heating, as seen in the age-velocity dispersion relation \citep[AVR;][]{1946ApJ...104...12S,1950AJ.....55..182R,1950ApJ...112..554R,1977A&A....60..263W,2004A&A...418..989N,2007A&A...475..519H,Holmberg2009,2007MNRAS.380.1348S,2008A&A...480...91S,2009MNRAS.397.1286A,Casagrande2011,Sharma2014,Yu2018,Raddi2022}.
Since the epoch of the \hipparcos\ mission \citep{1997A&A...323L..49P}, the kinematics of stars in the solar neighbourhood have been extensively studied \citep[e.g.][]{1998AJ....115.2384D,DB1998,1999A&AS..135....5C,1999MNRAS.308..731S,2004A&A...418..989N,2005A&A...430..165F,2008A&A...483..453F,2008A&A...490..135A}, and the precise astrometry of the \gaia\ mission \citep{2016A&A...595A...1G}, the successor to \hipparcos, has further advanced these studies \citep[e.g.][]{Bovy2017,2017A&A...608A..73K,2018A&A...616A..11G,Yu2018,2019MNRAS.484.3544R,Mikkola2022,Raddi2022}.

\gaia\ Early Data Release 3 \citep[EDR3;][]{2021A&A...649A...1G} has been used to study the kinematics of white dwarfs in the solar neighbourhood, showing that the kinematic structure of local white dwarfs \citep[e.g.][]{Mikkola2022} is similar to that of local main-sequence stars \citep[e.g.][]{2012MNRAS.426L...1A,2017A&A...608A..73K,2018A&A...616A..11G}.
For example, \citet{Raddi2022} performed a detailed three-dimensional kinematic analysis of a sample of 3133 white dwarfs from \gaia\ EDR3 with radial velocity measurements from either \gaia\ or cross-matched spectroscopic observations, finding that their sample, which was mostly located within $\sim 300$~pc of the Sun, consisted of $\sim 90-95$ per cent thin disc stars, $\sim 5-10$ per cent thick disc stars, and a few isolated white dwarfs and halo members. For the thin disc members, \citet{Raddi2022} determined their AVRs and found them to agree with previous AVR results that were determined from different samples of LAMOST-\gaia\ FGK-type stars without radial velocity information \citep{Yu2018}.

\gaia\ has also revealed structure in the colour-magnitude diagram of white dwarfs in the solar neighbourhood, which has implications for our understanding of white dwarf cooling.
\gaia\ Data Release 2 (DR2) revealed the so-called Q branch \citep{2018A&A...616A..10G}, an observed sequence of massive white dwarfs in the colour-magnitude diagram that is transversal to the theoretical cooling sequences for white dwarfs with mass $\gtrsim 1~M_\odot$ and coincides with the region of white dwarf crystallization \citep{Tremblay2019}. 
The phase transition from liquid to solid state during the process of crystallization results in the release of latent heat \citep{vanHorn1968} and other energy associated with element sedimentation \citep{1988A&A...193..141G,1994ApJ...434..641S,1997ApJ...485..308I,2012A&A...537A..33A} that slows the rate of white dwarf cooling, and this cooling delay results in a pile-up of white dwarfs that has been identified with the Q branch \citep{Tremblay2019}.

While a cooling delay due to white dwarf crystallization had been predicted for over 50 yr prior to \gaia\ DR2, \citet{Cheng2019} presented evidence from \gaia\ DR2 for an extra cooling delay of $\sim 8$~Gyr experienced by $\sim 6$ per cent of ultramassive  ($1.08-1.23~M_\odot$) white dwarfs on the Q branch that is not explained by standard crystallization models. 
By looking at the number density of white dwarfs as a function of the age inferred by white dwarf cooling models, \citet{Cheng2019} noted an excess number of ultramassive white dwarfs on the Q branch above the expected pile-up from the crystallization delay. 
Furthermore, \citet{Cheng2019} noted that for some ultramassive \gaia\ DR2 white dwarfs on the Q branch there was a discrepancy between the dynamic age, inferred from the transverse velocity using the AVRs of \citet{Holmberg2009} and \cite{Sharma2014} for the local thin and thick discs, and the photometric isochrone age, inferred from the photometry using white dwarf cooling models. Among $\sim 1.08-1.23~M_\odot$ white dwarfs, they found a population moving anomalously fast relative to the velocity expected from the AVR for their photometric ages.
\citet{Cheng2019} argued that this discrepancy could only be explained by an extra cooling delay, in addition to the delays due to crystallization and double white dwarf mergers, and suggested modifications to the treatment of $~^{22}\textrm{Ne}$ sedimentation in white dwarf cooling models as a possible mechanism.
This prompted many efforts to explain this cooling anomaly through modifications to white dwarf cooling models \citep{Bauer2020,Blouin2020,Blouin2021,Caplan2020,2021ApJ...919L..12C,2020PhRvD.102h3031H,2021ApJ...919...87B,Camisassa2021}.

In our recent work \citep[][hereafter \citetalias{massive}]{massive}, we used the improved \gaia\ EDR3 data to re-investigate this cooling anomaly by analysing the number density distribution of the photometric cooling ages for massive ($0.95-1.25~M_\odot$) white dwarfs identified in the white dwarf catalogue of \citet{GentileFusillo2021}. 
We considered a variety of publicly available white dwarf cooling models, including both the oxygen-neon core models of \citet{Camisassa2019} and carbon-oxygen core models of \citet{Bedard2020} with different envelope thicknesses. 
These are all standard cooling models, by which we mean that they do not attempt to reproduce the anomalous $\sim 8~\mathrm{Gyr}$ cooling delay proposed by \citet{Cheng2019}.
For each set of cooling models considered, the masses and ages of the white dwarfs were determined from the \gaia\ photometry, and the resultant cooling age distributions were compared to the expectation from the time-varying star formation rate observed by \citet{Mor2019} for \gaia\ DR2 main-sequence stars.
\citet{Mor2019} discovered a star formation burst in the local Galactic disc $2-3~\mathrm{Gyr}$ ago, which, if a uniform white dwarf birthrate were assumed, would appear as an excess number of massive white dwarfs around this age produced through single stellar evolution.
In \citetalias{massive}, it was indeed found that, under the assumption of a uniform birthrate, there was an apparent excess of white dwarfs both along the Q branch and below it, coinciding with the burst of star formation seen in the star formation history of main-sequence stars.

As part of the detailed analysis in \citetalias{massive}, the sample was further subdivided into three equally spaced mass bins, and the number density distribution of photometric cooling ages according to each set of cooling models was statistically compared to the \citet{Mor2019} star formation rate. 
For the two lightest mass bins, $0.95-1.05$ and $1.05-1.15~M_\odot$, it was found that standard cooling models could produce cooling age distributions that were statistically consistent with the expected distribution from the star formation rate observed by \citet{Mor2019}. 
For the most massive bin, $1.15-1.25~M_\odot$, it was found that the cooling age distribution was well fitted by a linear combination of the distribution expected for single stellar evolution products (taken to be proportional to the star formation rate) and the distribution expected for the product of double white dwarf merger products (taken to be the convolution of the star formation rate and the merger delay time distribution calculated by \citealp{Cheng2020} using population synthesis simulations) when approximately $40-50$ per cent of the $1.15-1.25~M_\odot$ white dwarfs that formed over the past 4~Gyr were produced through double white dwarf mergers. 
From this analysis, it was found in \citetalias{massive} that the photometric cooling age distribution of ultramassive white dwarfs could be explained by accounting for the time-dependent star formation rate and the presence of a large fraction of merger products among $1.15-1.25~M_\odot$ white dwarfs, without needing to invoke an anomalous $\sim 8~\mathrm{Gyr}$ cooling delay.

In this work, we follow up the work of \citetalias{massive} with a kinematic analysis of the transverse motions of the same sample of $0.95-1.25~M_\odot$ white dwarfs in the solar neighbourhood that was considered in \citetalias{massive}, subdivided into the same mass bins of $0.95-1.05$, $1.05-1.15$, and $1.15-1.25~M_\odot$ and with masses and cooling ages determined using white dwarf cooling models found in \citetalias{massive} to produce photometric cooling age distributions consistent with the expectation from the star formation history.
For each of these mass bins, we estimate the transverse velocity dispersion as a function of age using different estimators and analyse the distributions of the separate transverse velocity components for several age ranges.
We compare the empirical distributions to the expectation from the AVRs measured for main-sequence stars in the local thin disc and find a population of anomalously fast-moving ultramassive white dwarfs on the Q branch with kinematic features that are not explained by an extra cooling delay in white dwarfs originating from the local thin disc.

The structure of this paper is as follows.
The data and models used in this work, along with the formal procedure for determining transverse motions, are described in \cref{sec:methods}. 
The results are presented in \cref{sec:results}, and the implications for the origin of anomalously fast-moving ultramassive white dwarfs on the Q branch are discussed in \cref{sec:discussion}.
Finally, the results and discussion are summarized in \cref{sec:conclusions}.

\section{Methods} \label{sec:methods}

\subsection{Data Sample}

We use the data and white dwarf cooling models described in \citetalias{massive}. 
The data consist of \gaia\ EDR3\footnote{Although there is a more recent data release, \gaia\ Data Release 3 \citep{2022arXiv220800211G}, it does not contain new astrometric information compared to EDR3.} white dwarfs identified in the \citet{GentileFusillo2021} catalogue with at least 90 per cent probability, as determined by the \texttt{Pwd} parameter of the catalogue, and located within 200~pc of the Sun. 
The magnitude and colour observations were de-reddened following the prescription of \citet{GentileFusillo2021}.
We used extinctions of $A_G = 0.835 A_V$, $A_{G_\mathrm{BP}} = 1.139 A_V$, and $A_{G_\mathrm{RP}} = 0.650 A_V$ for the $G$-, $G_\mathrm{BP}$-, and $G_\mathrm{RP}$-band magnitudes, respectively, with $A_V$ values given by the \texttt{meanAV} parameter of the catalogue.
We calculated the absolute $G$-band magnitude, $M_G$, according to the expression
\begin{equation}
    M_G = G - 5 \log_{10}\left(\frac{100 \ \textrm{mas}}{\plx}\right) - A_G \ ,
\end{equation}
where $\plx$ is the parallax in $\textrm{mas}$, using the values of the catalogue parameters \texttt{parallax} and \texttt{phot\_g\_mean\_mag} for $\plx$ and $G$, respectively.

The completeness of this 200~pc sample as a function of $M_G$ was analysed in \citetalias{massive} and found to be volume limited for white dwarfs with $M_G \geq 15$. 
The limiting volume as a function of magnitude, $V_\mathrm{lim}(M_G)$, was determined in \citetalias{massive} using a variant of the \citet{1968ApJ...151..393S} estimator as follows. The cumulative number distribution as a function of sampling volume was constructed for magnitude-binned sub-samples, and $V_\mathrm{lim}$ was determined for each magnitude bin by finding the volume above which the distribution began to deviate from the linear relation expected for a complete sample (and which was realized at smaller volumes). 
The limiting volume for each white dwarf was then calculated by linearly interpolating between these magnitude-binned $V_\mathrm{lim}$ values as a function of $M_G$, taking the limiting volume for magnitudes $M_G \geq 15$ to simply be the total volume, $V_\mathrm{max}$, of the 200~pc sample (i.e. the volume of a sphere with a radius of 200~pc).

Following the procedure of \citetalias{massive}, we restrict our sample to only include white dwarfs within their completeness-limiting volumes and correct for the magnitude-dependent reduced sampling volume by assigning each remaining white dwarf a weight of
\begin{equation}
    w_i = 
    \begin{cases}
    V_\mathrm{max} \ / \ V_{\mathrm{lim},i} & \text{if} \ \ V_{\mathrm{lim},i} < V_\mathrm{max} \\
    1 & \text{otherwise}
    \end{cases} \ ,
\end{equation}
where $i$ is a data index denoting the $i$th white dwarf, $V_{\mathrm{lim},i}$ is the limiting volume for the magnitude of that white dwarf, and $w_i$ is the weight assigned to that white dwarf.

\subsection{Cooling and Atmosphere Models}

Using the publicly available \software{WD\_models} package provided by \citet{sihaocheng}\footnote{The \software{WD\_models} package is publicly available at \url{https://github.com/SihaoCheng/WD_models}}, we use white dwarf cooling models to determine the photometric age and mass of each white dwarf from its absolute $G$-band magnitude, $M_G$, and colour, $G_\mathrm{BP} - G_\mathrm{RP}$.
To determine masses and cooling ages from these photometry measurements, the atmosphere composition of the white dwarf is an important factor to consider.
As described in detail in \citetalias{massive}, we classified each source in our sample as having either a pure-H, pure-He, or mixed atmosphere based on the best-fitting results from \citet{GentileFusillo2021}, as determined by the chi-squared values provided in the catalogue for their model fits: \texttt{chisq\_H}, \texttt{chisq\_He}, and \texttt{chisq\_mixed}.
Each source was classified as having an atmosphere composition corresponding to whichever model gave the smallest chi-squared value. Sources for which all three chi-squared values were empty were assumed to have pure-H atmospheres.

Each set of models from \software{WD\_models} considered in \citetalias{massive} included both H-atmosphere and He-atmosphere models. 
The evolutionary models were combined with the corresponding publicly available Montreal group synthetic colours\footnote{The Montreal group synthetic colours are publicly available at \url{http://www.astro.umontreal.ca/~bergeron/CoolingModels}.} for pure-H and pure-He atmosphere models with \gaia\ EDR3 band-pass filters to convert the photometry measurements of each source into mass and cooling age values. 
The synthetic colours map the magnitude and colour to the effective temperature and surface gravity, respectively $T_\mathrm{eff}$ and $\log g$, and the evolutionary models map $T_\mathrm{eff}$ and $\log g$ to cooling age and mass. 
Models of the appropriate atmosphere composition were used for each source as determined by the atmosphere classification from the \citet{GentileFusillo2021} catalogue described above. 
He-atmosphere models were used for sources classified as having pure-He or mixed atmospheres, while H-atmosphere models were used for sources classified as having pure-H or unknown atmospheres.

In our previous work \citepalias{massive}, we assessed the consistency of the photometric cooling age distribution determined using various models available through \software{WD\_models} with the star formation history observed for \gaia\ DR2 main-sequence stars \citep{Mor2019}.
In this work, we use only the thick Montreal white dwarf cooling models \citep{Bedard2020}. These models gave the best-fitting cooling age distribution for the $1.05-1.15$ and $1.15-1.25~M_\odot$ mass bins in \citetalias{massive}. 
For the $0.95-1.05~M_\odot$ mass bin, the La Plata models \citep{Renedo2010,Camisassa2017,Camisassa2019} and thick Montreal models were both determined to produce cooling age distributions consistent with the star formation history and yielded similar \textit{p}-values, so for consistency we use the thick Montreal models for all three mass bins.
Using the La Plata models instead of the thick Montreal models for the lightest mass bin produces similar results.

Since the publication of \citetalias{massive}, the La Plata group has produced cooling models for ultramassive white dwarfs ($\gtrsim 1.15~M_\odot$) with carbon-oxygen cores \citep{Camisassa2022}. 
These models implement more realistic physics in the core of the white dwarfs than the \citet{Bedard2020} models, such as element diffusion and abundance ratios determined by progenitor evolution simulations, though there is still a great deal of uncertainty regarding the abundances expected for carbon-oxygen white dwarfs \citep[e.g.][]{Salaris2010,DeGeronimo2017,Wagstaff2020,Althaus2021}.
While an analysis of the photometric cooling age distributions determined by the \citet{Camisassa2022} models, particularly in comparison to the star formation rate, as was done in \citetalias{massive} is worthwhile, our choice of cooling model will not change the salient results presented in this work.
Using the \citet{Camisassa2022} models instead of the \citet{Bedard2020} models would shift the particular ages at which the key features of the velocity dispersion appear, but it would not remove these features.
Likewise, using oxygen-neon core models \citep[e.g.][]{Camisassa2019} instead of carbon-oxygen core models would shift the inferred cooling ages and, to a lesser extent, masses of our white dwarf sample, but it would not affect the values of the velocities and therefore would not eliminate the key features presented in this work.

\subsection{Kinematics}

\subsubsection{Streaming Motion}

The observed velocity of a star in the solar neighbourhood consists of the local (average) streaming velocity, which includes both the average motion due to Galactic rotation and reflex solar motion relative to the local standard of rest, plus the peculiar velocity of the star due to random motion. 
In this work, we are interested in the random motion of white dwarfs about the local mean, so we correct the observed velocities for both Galactic rotation and solar motion.

\citet{Olling2003} give a concise derivation of the local streaming motion in the Galactic plane to first order in a two-dimensional Taylor series expansion of the mean velocity field about the location of the Sun. To first order in this expansion, the Galactic rotation can be parametrized by the four Oort constants $A$, $B$, $C$, and $K$, with $C = K = 0$ in the case of axisymmetry.
The proper motions corresponding to the streaming velocity in Galactic coordinates are
\begin{align}
    \bar{\mu}_{\ell*} &=\left( U_0 \sin \ell - V_0 \cos \ell \right) \ \plx + \cos b \left(A \cos 2\ell - C \sin 2\ell + B \right)\\
    \begin{split}
        \bar{\mu}_b &= \left[\left( U_0 \cos \ell + V_0 \sin \ell \right) \sin b - W_0 \cos b \right] \ \plx\\
        &\qquad - \sin b \cos b \left( A \sin 2\ell + C \cos 2\ell + K \right) \ ,
    \end{split}
\end{align}
where the proper motions $\bar{\mu}_{\ell*}$ and $\bar{\mu}_b$ are given in units of $\textrm{km} \ \textrm{s}^{-1} \ \textrm{kpc}^{-1}$, $\bar{\mu}_{\ell*} \equiv \bar{\mu}_\ell \cos b$, $\plx$ is the inverse distance in units of $\textrm{kpc}^{-1}$ (corresponding to a parallax angle measured in $\textrm{mas}$), $\ell$ is the Galactic longitude, $b$ is the Galactic latitude, and $(U_0, V_0, W_0)$ is the velocity of the Sun relative to the local streaming in a heliocentric Cartesian coordinate system with $\uvec{x}$ pointing in the direction $(\ell, b) = (0^\circ, 90^\circ)$, $\uvec{y}$ in the direction $(\ell, b) = (90^\circ, 90^\circ)$, and $\uvec{z}$ in the direction of $b = 0^\circ$.
The components of the transverse streaming velocity in the $\uvec{\ell}$ and $\uvec{b}$ directions, in units of $\textrm{km} \ \textrm{s}^{-1}$, are 
\begin{equation}
    \left( \bar{v}_\ell, \ \bar{v}_b \right) = \left( \bar{\mu}_{\ell*}, \ \bar{\mu}_b \right) \ / \ \plx \ .
\end{equation}

We calculate the streaming velocities using values reported in the literature for both the solar motion and the Oort constants. For the solar motion, we use the peculiar solar velocity measurements of \citet{Wang2021}, with values of $U_0 = 11.69 \ \textrm{km} \ \textrm{s}^{-1}$, $V_0 = 10.16 \ \textrm{km} \ \textrm{s}^{-1}$, and $W_0 = 7.67 \ \textrm{km} \ \textrm{s}^{-1}$.
For the Galactic rotation, we assign the Oort constants the values measured by \citet{Bovy2017} using \gaia\ DR1 main-sequence stars from the Tycho-\gaia\ Astrometric Solution catalogue, with values of $A = 15.3 \ \textrm{km} \ \textrm{s}^{-1} \ \textrm{kpc}^{-1}$, $B = -11.9 \ \textrm{km} \ \textrm{s}^{-1} \ \textrm{kpc}^{-1}$, $C = -3.2 \ \textrm{km} \ \textrm{s}^{-1} \ \textrm{kpc}^{-1}$, and $K = -3.3 \ \textrm{km} \ \textrm{s}^{-1} \ \textrm{kpc}^{-1}$. 

A conversion factor of $\kappa = 4.74047$ is needed to convert proper motions given in $\textrm{mas} \ \textrm{yr}^{-1}$ to units of $\textrm{km} \ \textrm{s}^{-1} \ \textrm{kpc}^{-1}$. Including this conversion factor, the observed velocities (in $\textrm{km} \ \textrm{s}^{-1}$) in the $\uvec{\ell}$ and $\uvec{b}$ directions are
\begin{equation}
    \left( v_\ell^\textrm{(obs)}, \ v_b^\textrm{(obs)} \right) = \kappa \ \left( \mu_{\ell*}^\textrm{(obs)}, \ \mu_b^\textrm{(obs)} \right) \ / \ \plx \ .
\end{equation}
\gaia\ only provides proper motions in ICRS coordinates, so we convert the observed proper motions from ICRS coordinates to Galactic coordinates using the \software{astropy}\footnote{\url{http://www.astropy.org}} package for \software{Python} \citep{astropy:2013,astropy:2018}.

We calculate the peculiar velocities, $v_\ell$ and $v_b$, of the white dwarfs by correcting the observed velocities for solar motion and Galactic rotation
\begin{align}
    v_\ell &= v_\ell^\textrm{(obs)} - \bar{v}_\ell \label{eq:vl}\\
    v_b &= v_b^\textrm{(obs)} - \bar{v}_b \ . \label{eq:vb}
\end{align}
These two velocities, together with the sky coordinates and parallax, completely specify the random transverse motion of our sample relative to the local standard of rest.

\subsubsection{Random Transverse Motion}

Since most sources in \gaia\ EDR3 do not have radial velocity measurements, we perform our analysis of the kinematics using only the transverse motion.
Following the procedure of \citet{DB1998}, we relate the transverse motion to the full three-dimensional motion through use of a projection operator $\mat{P}$ that projects the three-dimensional space velocity $\vec{v}$ onto the celestial sphere. 

We write this formulation explicitly using matrix notation with vectors as column matrices and work in Cartesian Galactic coordinates in which $\uvec{x}$ points towards the Galactic centre ($\ell = 0^\circ$, $b = 90^\circ$), $\uvec{y}$ points in the direction of Galactic rotation ($\ell = 90^\circ$, $b = 90^\circ$), $\uvec{z}$ points towards the North Galactic Pole ($b = 0^\circ$), and $\vec{v} = \left(U, V, W\right)\transpose$.
Letting $\uvec{r} = \left( \cos b \sin\ell, \cos b \cos\ell, \sin b \right)\transpose$ be the unit position vector pointing to the location of the star, the projection operator in matrix notation is
\begin{equation}
    \mat{P} \equiv \mat{I} - \uvec{r} \, \uvec{r}\transpose \ ,
\end{equation}
where $\mat{I}$ is the $3 \times 3$ identity matrix.
The tangent velocity vector $\vec{v}_\perp$ is then related to $\vec{v}$ by
\begin{equation}
    \vec{v}_\perp = \mat{P} \, \vec{v} \ ,
\end{equation}
where $\vec{v}_\perp$ is the component of $\vec{v}$ in the tangent plane at the angular position of the star.

The rectangular components of the tangential velocity vector can be written explicitly as
\begin{equation}
\vec{v}_\perp = 
\begin{pmatrix}
U_\perp\\
V_\perp\\
W_\perp\\
\end{pmatrix}
= 
\begin{pmatrix}
- \sin \ell \ v_\ell - \cos \ell \ \sin b \ v_b \\
\hspace{1ex} \cos \ell \ v_\ell - \sin \ell \ \sin b \ v_b \\
\cos b \ v_b
\end{pmatrix} \ ,
\end{equation}
and the magnitude of the transverse velocity, $v_t =| \vec{v}_\perp|$, is simply given by
\begin{equation}
v_t = \sqrt{v_\ell^2 + v_b^2} \ .
\end{equation}
We calculate $v_t$ and the components of $\vec{v}_\perp$ for the white dwarfs in our sample using $v_\ell$ and $v_b$ as given by \cref{eq:vl,eq:vb}.
Since the mean local motion due to both Galactic rotation and solar motion has already been removed, the velocities here are the peculiar velocities, corresponding to the random motion about the mean.

If the full three-dimensional space velocities (including radial motion) have a diagonal dispersion tensor $\mat{\sigma}^2 = \mathrm{diag}\left(\sigma_U^2, \sigma_V^2, \sigma_W^2 \right)$ and the positions of the white dwarfs are uncorrelated with the velocities, then the dispersion of the transverse velocity can be quantified by
\begin{equation}
    \sigma_t^2 = E\left[ \mathrm{Tr}\left( \mat{P} \mat{\sigma}^2 \mat{P}\transpose \right) \right] \ , \label{eq:sigma_t}
\end{equation}
where $\mathrm{Tr}$ denotes the matrix trace, which is equivalent to the sum of eigenvalues of the matrix, and $E$ denotes the expectation value for a particular distribution of sky coordinates, which can be estimated by a statistic such as the sample mean.

Throughout this work, we compare the results we find for the white dwarf transverse kinematics relations with the expectation assuming the dispersions of their full three-dimensional velocities follow the AVRs determined by \citet{Holmberg2009} for main-sequence stars from the local thin disc. While more recent determinations of these AVRs exist \citep[e.g.][]{Yu2018}, they are similar to the AVRs of \citet{Holmberg2009} and do not change our results. 
As evidence for a cooling anomaly among ultramassive white dwarfs was presented by \citet{Cheng2019} with reference to the thin disc AVRs of \citet{Holmberg2009}, using the \citet{Holmberg2009} AVRs facilitates comparison of our results with earlier work on this topic.

The sample for which the \citet{Holmberg2009} AVRs were determined was volume-complete to a distance of $\sim 40~\mathrm{pc}$ \citep{2004A&A...418..989N,2007A&A...475..519H,Holmberg2009}, so both the \citet{Holmberg2009} sample and our $200~\mathrm{pc}$ sample are local enough that bias against populations with large scale height is not a major concern in comparing our results to the \citet{Holmberg2009} AVRs. 
This was verified by comparing the \citet{Holmberg2009} AVRs to the \citet{Yu2018} AVRs for a sample of
thin disc (metal-rich) stars with $|z| < 270~\textrm{pc}$, which we found to be similar enough that the main features in the comparison of our results to the AVRs were unchanged. 
The similarity of the \citet{Yu2018} AVRs to the \citet{Holmberg2009} AVRs demonstrates the negligible impact of scale height bias for the results presented in this work.

\section{Results} \label{sec:results}

\subsection{Summary Statistics} \label{ssec:res_summary_stats}

\begin{figure}
    \centering
    \includegraphics[width=\columnwidth,clip,trim=0.15in 0.15in 0.15in 0.15in]{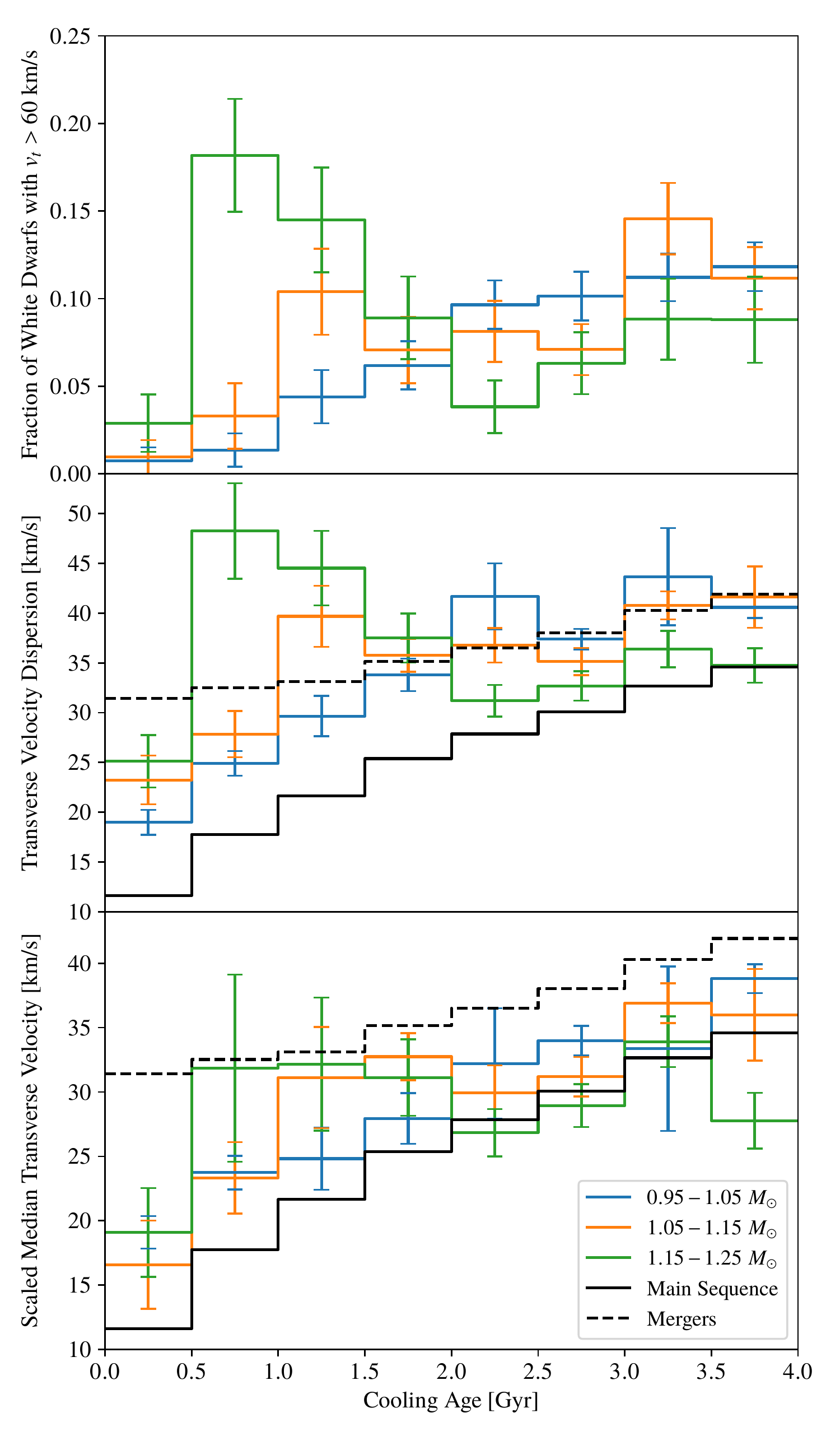}
    \caption{Upper panel: Fraction of fast-moving massive white dwarfs for three mass bins. 
    The fast-moving white dwarfs are defined as those with transverse velocity greater than $60~\mathrm{km}~\mathrm{s}^{-1}$. 
    Middle panel: Transverse velocity dispersion accounting for uncertainties in proper motion. 
    Lower panel: Scaled median transverse velocity (robust estimator of velocity dispersion).
    In the middle and lower panels, the solid black line traces the results for main-sequence stars from \citet{Holmberg2009} with the same sky distribution as the white dwarfs, and the dashed black lines trace the results from main-sequence stars including a merger delay described by the merger delay distribution of \citet{Cheng2020}.
    The main-sequence curve (solid black) is the expectation if the $1.15-1.25~M_\odot$ white dwarfs are all produced by single progenitors, while the mergers curve (dashed black) is the expectation in the limit that 100 per cent of those white dwarfs are produced by double white dwarf mergers.}
    \label{fig:umassive_disp}
\end{figure}

\cref{fig:umassive_disp} shows various probes of the AVR for \gaia\ EDR3 white dwarfs in three different mass bins, $0.95-1.05~M_\odot$ (blue), $1.05-1.15~M_\odot$ (orange), and $1.15-1.25~M_\odot$ (green). 
The masses and cooling ages were calculated from the observed photometry using the single stellar evolution white dwarf cooling models of \citet{Bedard2020} which were determined in \citetalias{massive} to produce cooling age distributions that are statistically consistent with the star formation rate observed for \gaia\ DR2 main-sequence stars \citep{Mor2019}.

The upper panel of \cref{fig:umassive_disp} shows the fraction of fast-moving white dwarfs with $v_t > 60 \ \mathrm{km} \ \mathrm{s}^{-1}$ as a function of photometric cooling age. 
The empirical distributions, shown as histograms, were constructed using cooling age bins of $0.5~\mathrm{Gyr}$ width over the range $0-4~\mathrm{Gyr}$. 
The fraction of fast-movers in each age bin was calculated using weighted counts for that bin, where each white dwarf was assigned a weight according to the procedure of \citetalias{massive} to correct for the reduced sampling volume that ensures completeness.

The fraction of fast-moving ultramassive white dwarfs in \gaia\ EDR3 as a function of photometric cooling age shows a trend with cooling age that is similar to what \citet{Cheng2019} found for \gaia\ DR2 white dwarfs, though we see a less pronounced peak for young $1.05-1.15~M_\odot$ white dwarfs on the Q branch than what was found by \citet{Cheng2019} for this bin.
For the $0.95-1.05~M_\odot$ white dwarfs, the fraction of fast-movers tends to increase with cooling age, which is the expected behaviour if the white dwarf velocities follow an AVR similar to what has been observed for main-sequence (G and F dwarf) stars in the thin disc \citep{Holmberg2009}.
Of particular note is the distribution for $1.15-1.25~M_\odot$ white dwarfs, for which the fraction of young fast-movers in the age range $0.5-1.5~\mathrm{Gyr}$ is larger than the fraction in either the younger or older age bins. 
This corresponds to an excess of young fast-movers on the Q branch relative to the expectation based on the AVR observed for stars in the thin disc \citep{Holmberg2009}.
While we see a hint of a similar excess in the $1.5-2.0~\mathrm{Gyr}$ bin of the $1.05-1.15~M_\odot$ white dwarfs, we do not see a clear signal of an excess of fast-movers within the uncertainties. 
This difference between our distribution and that of \citet{Cheng2019} for the $1.05-1.15~M_\odot$ white dwarfs could be due to the difference in colours and proper motions between the \gaia\ EDR3 and DR2 measurements. 
The updated colour measurements for very blue objects in particular affect the masses inferred for those objects.

The middle and lower panels of \cref{fig:umassive_disp} show estimates of the transverse velocity dispersion as a function of age using different estimators of $\sigma_t$, one based on the sample mean in each bin (middle panel) and the other based on the sample median (lower panel), the latter of which is more robust to outliers.
Random variation in the observed values of $v_t$ will arise both from the intrinsic randomness in the motion of the white dwarfs about the local mean (i.e. the peculiar motion of the white dwarfs) that we are interested in and from random measurement error. We correct for the latter effect in both methods of estimating $\sigma_t$ from the observations. 
The random error for each value of $v_t$ was calculated by propagating the standard errors and correlations provided by \gaia\ for the proper motions and parallax from which the observed transverse velocity (without corrections for Galactic rotation or solar motion) was calculated. 
The error was propagated using the expression $\evto^2 = \mat{J} \, \matsymb{\Sigma} \, \mat{J}\transpose$, where $\matsymb{\Sigma}$ is the covariance matrix, $\mat{J}$ is the Jacobian matrix, and $\evto$ is the random error for the observed transverse velocity.

The middle panel of \cref{fig:umassive_disp} shows the transverse velocity dispersion calculated using the mean for each age (and mass) bin with a correction for random measurement error according to the expression
\begin{equation}
    \sqrt{\Bigl\langle v_t^2 \Bigr\rangle - \Bigl\langle \evto^2 \Bigr\rangle} \ ,
\end{equation}
where $v_t$ is the transverse velocity after correcting for Galactic rotation and reflex solar motion (i.e. calculated from the sky-projected velocity after subtracting the mean velocity), $\evto$ is the standard error for the transverse velocity calculated from observed proper motions and parallax without subtracting the mean Galactic rotation and reflex solar motion, and the angle brackets $\left\langle~\right\rangle$ denote a weighted sample average (e.g. $\left\langle v_t^2 \right\rangle = (\sum_i w_i v_t^2) / (\sum_i w_i)$ with $i$ an index labeling the data points in the sample).

The lower panel of \cref{fig:umassive_disp} shows the scaled median velocity corrected for random measurement error, which is calculated for each joint mass and photometric cooling age bin using the formula
\begin{equation}
    \sqrt{\frac{\widetilde{v}_t^2}{\ln 2} - \left\langle \evto^2 \right\rangle} \ ,
\end{equation}
where $\widetilde{v}_t$ is the sample median of $v_t$, determined for each bin using the weighted cumulative fraction for white dwarfs in that bin.
The motivation for scaling the median is that we empirically find the the transverse velocities to be approximately Rayleigh-distributed. A Rayleigh distribution parametrized by mode $\sigma$ has a mean of $\sigma \sqrt{\pi/2}$, median of $\sigma \sqrt{2 \ln 2}$, and variance of $\sigma^2 (2-\pi/2)$, yielding a transverse velocity dispersion of $\sigma\sqrt{2}$ and $ \widetilde{v}_t^2 = (\ln 2 )\langle v_t^2 \rangle$.

In both the middle and lower panels of \cref{fig:umassive_disp}, the solid black line shows the expected relation according to the AVRs of main-sequence (G and F dwarf) stars found by \citet{Holmberg2009} for the sky distribution of the sample of $1.15-1.25~M_\odot$ white dwarfs in each age bin.
This curve assumes that all the white dwarfs formed through single stellar evolution, and thus that none of the white dwarfs formed through mergers.
The transverse velocity dispersion value for each bin was calculated by taking the sample average of the sum of the eigenvalues of $\mat{P} \mat{\sigma}^2 \mat{P}\transpose$ (equivalent to the trace of this matrix),
\begin{equation}
    \left\langle \textrm{Tr}\left( \mat{P} \mat{\sigma}^2 \mat{P}\transpose \right) \right\rangle \ ,
\end{equation}
where $\mat{P}$ is the projection operator that projects the velocity $\vec{v}$ onto the tangent plane of the celestial sphere such that $\vec{v}_\perp = \mat{P} \, \vec{v}$, $v_t = |\vec{v}_\perp|$, and $\mat{\sigma}^2 = \textrm{diag}\left(\sigma_U^2, \sigma_V^2, \sigma_W^2 \right)$ is the dispersion tensor of the full three-dimensional space velocities before projection. The values of $\sigma_U$, $\sigma_V$, and $\sigma_W$ are calculated for each white dwarf using the AVRs determined by \citet{Holmberg2009} with the photometric cooling age of that white dwarf.
The AVRs of \citet{Holmberg2009} for the three velocity components and the total velocity (giving $\sigma_U$, $\sigma_V$, $\sigma_W$, and $\sigma_\mathrm{tot}$ as a function of age) each follow a power law with exponents of 0.39, 0.40, 0.53, and 0.40 for $U$, $V$, $W$, and total dispersion, respectively.

We similarly determined the expected relation for the \citet{Holmberg2009} AVRs if all of the white dwarfs are the product of double white dwarf mergers, which is shown as the dashed black curve in the middle and lower panels of \cref{fig:umassive_disp}. 
This corresponds to the limit in which the merger fraction is 100 per cent for white dwarfs that formed over the last 4~Gyr.
Instead of using the photometric cooling age inferred directly using the white cooling models applicable for single stellar evolution to calculate the predicted $\sigma_U$, $\sigma_V$, and $\sigma_W$ values from the AVRs, the relation for merger products was calculated using the expected true age after accounting for a merger delay time, with the expectation value calculated using the merger delay distribution of \citet{Cheng2020}. 
Delay time distributions associated with double white dwarf mergers were calculated by \citet{Cheng2020} using binary population synthesis simulations for resultant white dwarfs of various mass ranges; we use the distribution for $1.14-1.24~M_\odot$ white dwarfs, which spans approximately the same mass range as our most massive bin and is the merger delay time distribution that was used in \citetalias{massive}.

Double white dwarf mergers can produce photometrically young white dwarfs with transverse velocities faster than what would be expected for single stellar evolution products of the same photometric age because the true age of the white dwarf merger products (accounting for the lifetime of the merger progenitors) is typically older than the age inferred from photometry \citep{Temmink2020}. 
Estimates of the merger fraction from both simulations \citep[e.g.][]{Bogomazov2009,Temmink2020} and data \citep[e.g.][]{Cheng2020,Kilic2021,2023MNRAS.518.2341K,massive} indicate that double white dwarf merger products constitute an appreciable fraction of ultramassive white dwarfs, with estimates ranging from $\sim 15$ per cent to over 50 per cent. 
The population synthesis simulations of \citet{Temmink2020} indicate that $30-50$ per cent of white dwarfs with mass $\geq 0.9~M_\odot$ form through binary mergers, with about 45 per cent of these mergers occurring between two white dwarfs, while the population synthesis simulations of \citet{Bogomazov2009} indicate that over 50 per cent of white dwarfs with mass $\geq 1.1~M_\odot$ are produced through double white dwarf mergers.

Empirically, based on the magnetism, kinematics, rotation, and composition of the 25 most massive white dwarfs identified by \citet{Kilic2021} in the Montreal White Dwarf Database 100~pc sample, \citet{2023MNRAS.518.2341K} found a merger fraction of $56^{+9}_{-10}$ per cent among ultramassive white dwarfs with mass $\gtrsim 1.3~M_\odot$. 
Theoretical results indicate that nearly half of the identified merger products would have been produced through the double white dwarf merger channel \citep{Temmink2020}, and over half of the merger products identified in \citet{Kilic2021,2023MNRAS.518.2341K} showed kinematic and/or rotation features indicative of double white dwarf mergers in particular. 
\citet{Cheng2020} estimated a double white dwarf merger fraction of about 20 per cent for \gaia\ DR2 white dwarfs within 250~pc with mass in the range $0.8-1.3~M_\odot$ and a larger fraction of about 35 per cent for the subset in the mass range $1.14-1.24~M_\odot$ (assuming carbon-oxygen cores to determine the mass). 
For our \gaia\ EDR3 white dwarf sample within 200~pc in the $1.15-1.25~M_\odot$ mass bin, it was found in \citetalias{massive} that the photometric cooling age distribution was well fitted when the double white dwarf merger fraction was about $40 - 50$ per cent.

The expectation for the limiting case of 100 per cent double white dwarf merger products is shown in \cref{fig:umassive_disp} as a simple check of the feasibility of a double white dwarf merger delay explanation of the fast-movers. 
The lower panel of \cref{fig:umassive_disp} shows that a sufficiently large fraction of merger products among $1.15-1.25~M_\odot$ white dwarfs can reach the excess observed for the median estimate of the transverse velocity dispersion, which is representative of the (more slowly moving) bulk of the white dwarfs as the median estimator is robust to the outlying fast-movers. 
However, even in the extremum limit that 100 per cent of the $1.15-1.25~M_\odot$ white dwarfs are the product of double white dwarf mergers, the middle panel of \cref{fig:umassive_disp} shows that the excess observed for the mean estimate of the transverse velocity dispersion, which is sensitive to the outlying fast-movers, cannot be achieved. 
In keeping with \citet{Cheng2019}, we thus find that double white dwarf mergers (in the local Galactic disc) cannot be the sole explanation of the fast-movers.

\subsection{Three-Dimensional Velocity Distributions} \label{ssec:res_3d_dists}

\begin{figure*}
    \centering
    \includegraphics[width=\textwidth,clip,trim=0.1in 0in 0.1in 0.1in]{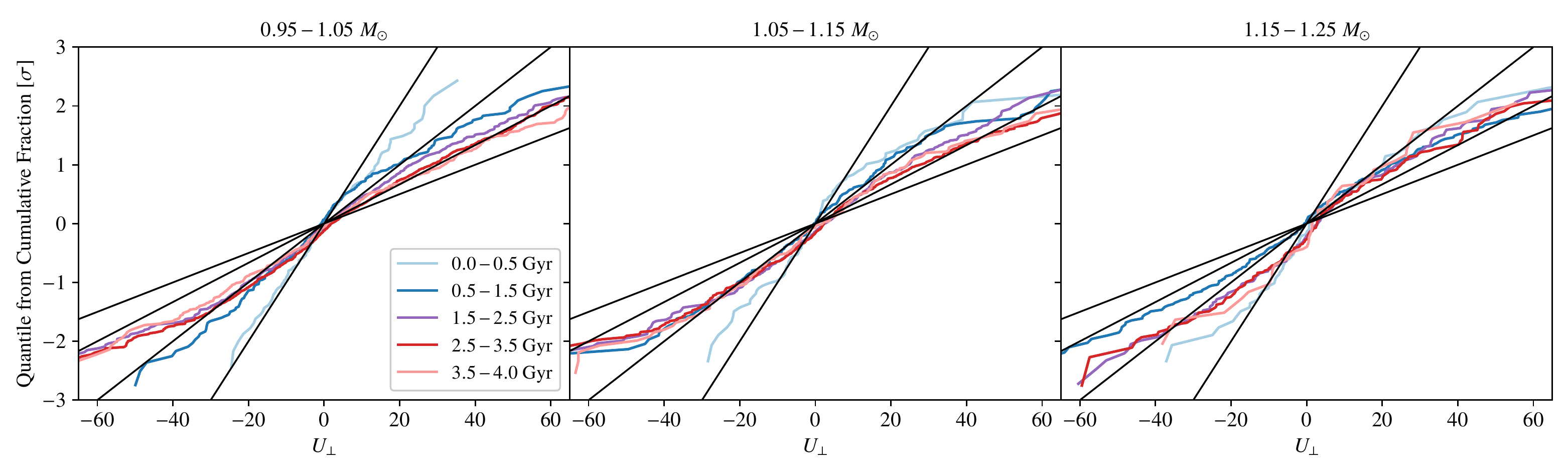}\\
    \includegraphics[width=\textwidth,clip,trim=0.1in 0in 0.1in 0in]{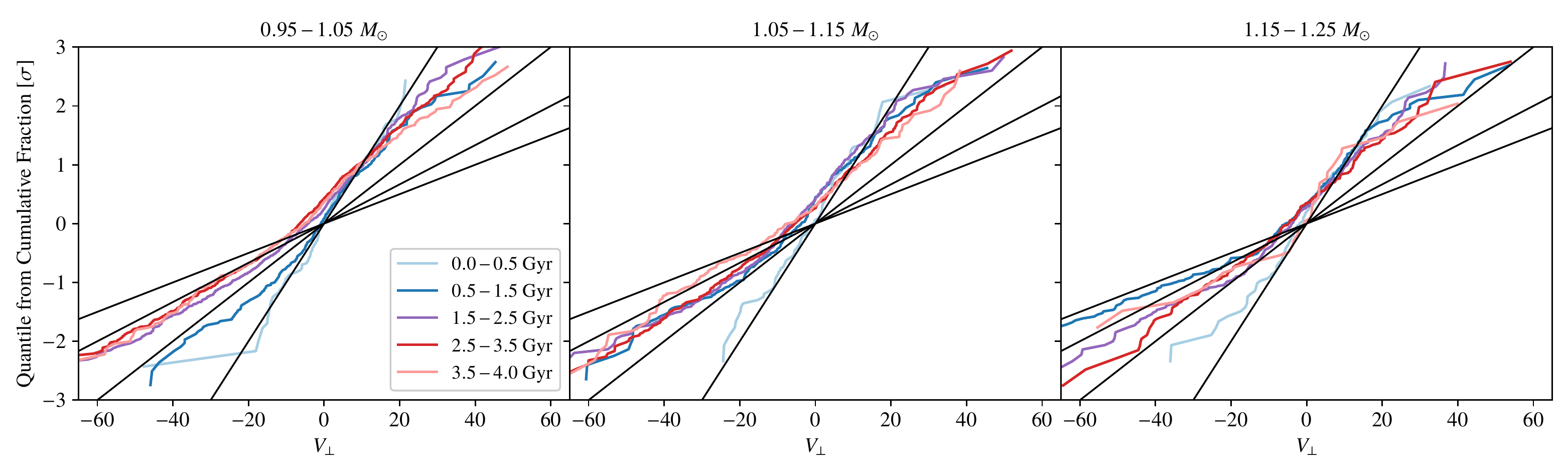}\\
    \includegraphics[width=\textwidth,clip,trim=0.1in 0.1in 0.1in 0in]{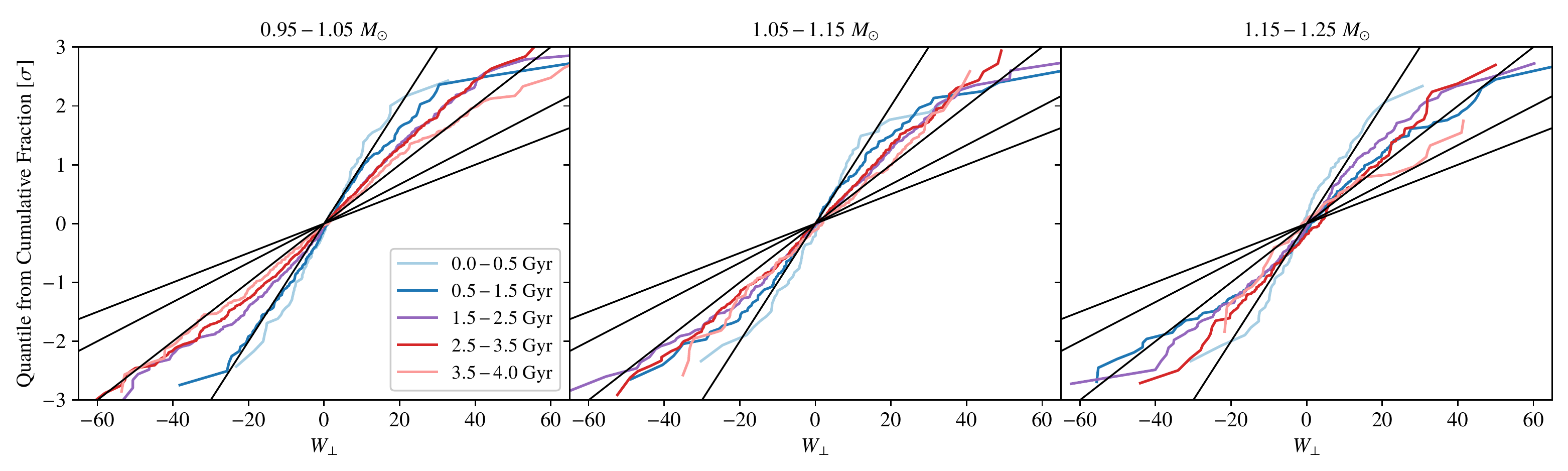}\\
    \caption{Quantile-quantile plots of transverse velocity components.
    The vertical axis is the quantile function in units of standard deviation $\sigma$ for a normal distribution with zero mean, calculated from the empirical cumulative fraction corresponding to the velocity component shown on the horizontal axis.
    The black lines correspond to normal distributions with $\sigma = 10, 20, 30$, and $40~\mathrm{km}~\mathrm{s}^{-1}$ in order of decreasing slope.}
    \label{fig:uvw_quantiles}
\end{figure*}

The behaviour of the AVRs that we measure in \cref{ssec:res_summary_stats} for the transverse velocities aligns with the behaviour of the observed fraction of fast-movers as a function of cooling age.
For both the non-robust and robust estimators of transverse velocity dispersion, we see that the lightest ($0.95-1.05~M_\odot$) white dwarfs tend to become more dispersed with age, as inferred from the fraction of fast-movers and expected from the AVRs of main-sequence stars. 
The $1.05-1.15~M_\odot$ white dwarfs mostly follow this trend as well, but with a small peak for cooling ages of $\sim 1.0-2.0$~Gyr. The most massive ($1.15-1.25~M_\odot$) white dwarfs show a prominent peak for cooling ages of $\sim 0.5-2.0$~Gyr.
The peak in transverse velocity dispersion of ultramassive white dwarfs inferred using the non-robust estimator (mean) is much larger than that inferred using the robust estimator (median), suggesting that the fast-movers in these age bins are outliers or comprise a separate population that is much more dispersed than the main population. 

To better understand the kinematic features of the population of young fast-movers, we consider the distribution of each Cartesian component of the transverse velocities in each mass and age bin.
\cref{fig:uvw_quantiles} shows quantile-quantile (Q-Q) plots comparing the sample distribution of transverse Cartesian velocity components $U_\perp$ (upper row), $V_\perp$ (middle row), and $W_\perp$ (bottom row) to the quantile function for a normal distribution with zero mean.
The quantile function of a distribution is the inverse of the cumulative distribution; for an input percentile $p$, it gives the value of the random variable for which there is $p$ probability of the random variable being less than or equal to the output value.
For a 1D normal distribution with mean $\mu$ and variance $\sigma^2$, the quantile function is $\mu + \sigma \ \sqrt{2} \ \mathrm{erfc}^{-1}\left( 2 \left[1 - p \right] \right)$, where $p \in [0,1]$ is the percentile (i.e. cumulative probability) and $\mathrm{erfc}^{-1}$ is the inverse complimentary error function.
In \cref{fig:uvw_quantiles}, the measured value of the velocity component is used as the $x$-axis variable, while the $y$-axis variable is the value of the quantile function in units of standard deviation calculated from the weighted cumulative fraction. 
Written explicitly, the $y$-axis variable is $y_i = \sqrt{2} \ \mathrm{erfc}^{-1}\left( 2 \left[1 - p_i \right] \right)$ for the $i$th ordered data point (in order of increasing $x$) with $p_i = \sum_{j\leq i} w_j$ where $j$ runs over all points for which $x_j \leq x_i$.
For reference, we also show quantile functions for normal distributions with zero mean and $\sigma = 10, 20, 30$, and $40~\mathrm{km}~\mathrm{s}^{-1}$ in order of decreasing slope as black lines.
If the transverse velocity in a particular direction is normally distributed for a particular sample, the corresponding Q-Q plot in \cref{fig:uvw_quantiles} should appear as a straight line. 
A non-zero mean will simply cause a shift of the distribution along the horizontal axis, as can be seen in the plots for $V_\perp$ due to asymmetric drift.

Observations of main-sequence stars \citep[e.g.][]{Holmberg2009} indicate that the velocity dispersion of stars in the local thin disc increases with age. 
If the white dwarf velocities follow the AVRs of main-sequence stars \citep{Holmberg2009}, the slope of the distribution in \cref{fig:uvw_quantiles} should get progressively less steep with increasing cooling age for a given mass bin and velocity component. 
For the age bins considered in \cref{fig:uvw_quantiles}, this effect should be more pronounced for younger ages due to the power-law form of the AVRs with exponents $< 1$. 
This is the approximate behaviour seen for the lightest two bins, $0.95-1.05~M_\odot$ and $1.05-1.15~M_\odot$ (left and middle columns of \cref{fig:uvw_quantiles}).
For the ultramassive $1.15-1.25~M_\odot$ white dwarfs, however, we see that the curves for the $0.5-1.5$~Gyr white dwarfs have a shallower slope and thus larger dispersion than what would be predicted by the \citet{Holmberg2009} AVRs, particularly for negative $U_\perp$ and $V_\perp$.
The dispersions for these $0.5-1.5$~Gyr white dwarfs are notably larger than the dispersions for the older $1.5-2.5$ and $2.5-3.5$~Gyr white dwarfs with mass in the range $1.15-1.25~M_\odot$.
Furthermore, for the $0.5-1.5$~Gyr age bin of $1.15-1.25~M_\odot$ white dwarfs, the white dwarfs with negative $V_\perp$ are much more dispersed in $V_\perp$ than in $U_\perp$, indicating that these white dwarfs likely do not originate from the local thin disc. 

\begin{figure}
    \centering
    \includegraphics[width=\columnwidth,clip,trim=0.15in 0.15in 0.15in 0.15in]{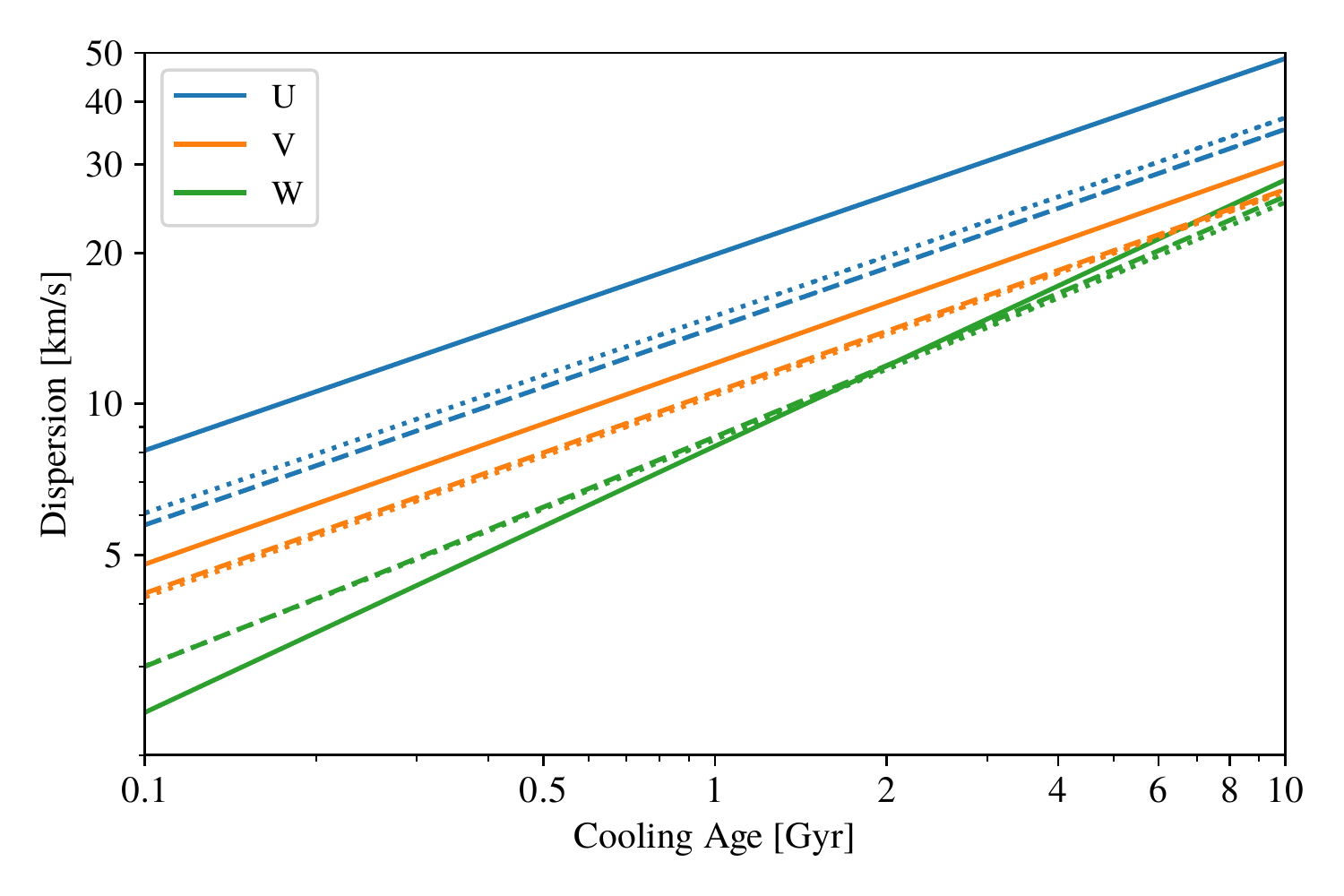}
    \caption{AVRs of projected and un-projected velocities. 
    AVRs are shown for $U$, $V$, and $W$ in blue, orange, and green, respectively. 
    The solid curves show the AVRs for the full three-dimensional space velocities determined by \citet{Holmberg2009} for main-sequence stars. 
    The dotted curves show the expected AVRs for the projected velocities assuming that the un-projected velocities follow the AVRs of \citet{Holmberg2009} and that the sky distribution is isotropic. 
    The dashed curves also show the expected AVRs for the \citet{Holmberg2009} projected velocities but assuming that the sky distribution is that observed for $1.15-1.25~M_\odot$ white dwarfs with $0.5-2.5$~Gyr cooling ages.}
    \label{fig:AVRs}
\end{figure}

The AVRs of \citet{Holmberg2009} indicate that local thin disc stars are more dispersed in $U$ than $V$, and we show in \cref{fig:AVRs} that this expectation holds for the tangential velocity components $U_\perp$ and $V_\perp$ as well.
Since the \citet{Holmberg2009} AVRs describe the full three-dimensional space velocities, we converted those relations to the expected AVRs of the projected velocities for the sky coordinate distribution of our $1.15-1.25~M_\odot$ sample with $0.5-2.5$~Gyr photometric ages.
This enables us to properly compare the results of \citet{Holmberg2009} for main-sequence stars from the local thin disc to the dispersion values of the tangential velocity components inferred from \cref{fig:uvw_quantiles} for our ultramassive white dwarf sample.
The expected AVRs of the projected velocities are given by the square root of the diagonal elements of the expectation of the matrix $\mat{P} \mat{\sigma}^2 \mat{P}\transpose$. 
\cref{fig:AVRs} shows the AVRs from \citet{Holmberg2009} for the full three-dimensional space velocities for each of the U, V, and W components (as solid blue, orange, and green curves, respectively), compared to the computed AVRs expected from the projected velocities for our sample's empirical sky distribution (dashed lines).
We also analytically evaluated the expected AVRs of the projected velocities for an isotropic distribution (shown as the dotted curves in \cref{fig:AVRs}) and found them to be nearly the same as the AVRs determined using the empirical sky distribution of our sample. 

From \cref{fig:AVRs} we see that the expected dispersions for the projected velocities are typically notably smaller than the corresponding dispersion of the un-projected velocities, but that the relative size of the dispersion for $V_\perp$ compared to $U_\perp$ is similar to the relative sizes of $\sigma_V$ and $\sigma_U$. 
Of particular note is that, throughout the age of the thin disc, the dispersion of $V_\perp$ is smaller than the dispersion of $U_\perp$.
For the population of anomalously fast white dwarfs (mass bin $1.15-1.25~M_\odot$ and photometric age bin $0.5-1.5$~Gyr) with negative $V_\perp$, however, the dispersion of $V_\perp$ is actually larger than the dispersion of $U_\perp$.
A population in equilibrium in the disc should have a dispersion ratio for $V$ relative to $U$ of $\sigma_V / \sigma_U = 2/3$ \citep{2008gady.book.....B}, and the AVRs measured by \citet{Holmberg2009} for thin disc main-sequence stars have a dispersion ratio nearly equal to this equilibrium ratio (though actually slightly smaller). The dispersion of $V_\perp$ being larger than the dispersion of $U_\perp$ in the anomalous population thus indicates that this population is not in equilibrium in the disc and thus does not originate from the local thin (or thick) disc.

\section{Discussion} \label{sec:discussion}

The presence of a population of anomalously fast-moving young, ultramassive white dwarfs on the Q branch was first discovered in \gaia\ DR2 observations by \citet{Cheng2019}, who argued this population must experience an extra cooling delay on the Q branch of $\sim 6 - 8$~Gyr that is not accounted for in current white dwarf cooling models.
This proposed extra cooling delay scenario explained the large transverse velocities observed for the fast-moving population as a consequence of these white dwarfs being much older than their photometric ages indicate. The fast-movers in this scenario were assumed to originate from the local disc, where observed AVRs indicate that objects become more dispersed over time. 
As a check of the assumption of a local disc origin of the fast-mover population in the extra cooling delay scenario, \citet{Cheng2019} ran a test in which they modelled the velocity distribution of the fast-movers on the Q branch as a Gaussian and determined kinematic relations expected for a disc in equilibrium, notably finding a dispersion ratio $\sigma_V/\sigma_U$ of approximately $2/3$. 
Like \citet{Cheng2019}, we also find a population of anomalously fast-moving young, ultramassive white dwarfs coincident with the Q branch.
However, the empirical distributions of the transverse velocity components in \cref{fig:uvw_quantiles} for \gaia\ EDR3 reveal kinematic features that are inconsistent with a local disc origin, in particular the distribution for $V_\perp$ and the dispersion ratio for $V_\perp$ and $U_\perp$.

\cref{fig:uvw_quantiles} suggests that the distribution of $V_\perp$ for the anomalously fast ultramassive white dwarfs is actually too broad to be explained entirely by a cooling delay in (thin) disc white dwarfs.
If white dwarfs with a true age of 10~Gyr produced through single stellar evolution in the Galactic thin disc follow the AVRs determined by \citet{Holmberg2009}, then the expected dispersion for the projected velocity component $V_\perp$ in the tangent plane averaged over the sky distribution of the $1.15-1.25~M_\odot$ white dwarfs with photometric ages $0.5-2.5$~Gyr would be approximately $26~\mathrm{km}~\mathrm{s}^{-1}$ (see \cref{fig:AVRs}).
From \cref{fig:uvw_quantiles}, we see that the anomalous $1.15-1.25~M_\odot$ white dwarfs with $0.5-1.5$~Gyr photometric ages have a dispersion for $V_\perp$ of $\sim 40~\mathrm{km}~\mathrm{s}^{-1}$ when $V_\perp$ is negative (i.e. among white dwarfs that are moving in the opposite direction of Galactic rotation). The Galactic thin disc is not old enough to produce white dwarfs with such a high dispersion given the observed AVRs \citep{Holmberg2009}.
While stars in the older thick disc have slightly higher dispersion \citep[e.g.][]{Sharma2014}, the expected dispersion of $\sim 33~\mathrm{km}~\mathrm{s}^{-1}$ for $V_\perp$ is still too small to explain this dispersion through a cooling delay in a thick disc population.

Furthermore, as noted in \cref{ssec:res_3d_dists} and seen in \cref{fig:uvw_quantiles}, we find a population of highly dispersed ultramassive white dwarfs coincident with the Q branch for which the dispersion in $V_\perp$ is larger than the dispersion in $U_\perp$. This is inconsistent with the dispersion ratio of a disc in equilibrium, for which the dispersion in $V_\perp$ should be smaller than the dispersion in $U_\perp$ with a dispersion ratio similar to that of the un-projected velocity components (see \cref{fig:AVRs}). 
Thus the empirical dispersion ratio for $V_\perp$ and $U_\perp$ suggests that these white dwarfs do not originate from the Milky Way disc.

The true origin of this population, however, is unclear. 
These white dwarfs may have originated from elsewhere in the Galaxy such as from the halo. 
The halo is older than the Galactic disc \citep[e.g.][]{2019NatAs...3..932G} and halo stars are typically more dispersed \citep{2018A&A...616A..11G}.
This scenario still requires a mechanism by which photometrically young white dwarfs acquire velocities faster than their photometric age would indicate. 
This mechanism could be an extra cooling delay as proposed by \citet{Cheng2019} of nearly 10~Gyr, so that only white dwarfs with photometric ages of $0.5-1.5$~Gyr are affected.  Alternatively, these objects could originate from a dynamically distinct, hot yet young population.

Another possible explanation of the anomalous fast-movers is that they are the merger remnant of the inner binary of a hierarchical triple system whose outer companion was ejected.
Observations indicate that triple systems are common, constituting about 10 per cent of multiples with F and G dwarf primaries \citep{2008MNRAS.389..869E,2010ApJS..190....1R,2014AJ....147...87T,2017ApJS..230...15M} and much larger fractions for higher mass B- and O-type primaries \citep{2014ApJS..215...15S,2017ApJS..230...15M}, and population synthesis simulations indicate that stellar interactions occur in the majority of triple systems \citep{2020A&A...640A..16T,2022MNRAS.516.1406S}.
The evolution of triple systems is significantly more complicated than that of isolated binaries due to the combination of stellar evolution, stellar interactions, and three-body dynamics \citetext{see e.g. \citealp{2016ComAC...3....6T} for a review}.
Triple systems tend to be hierarchical, with an inner binary and a distant companion \citep{1983ApJ...268..319H}.
Such systems have a large variety of possible evolution channels \citep{2020A&A...640A..16T,2022MNRAS.516.1406S,2022A&A...661A..61T} and can result in mergers and ejections \citep{2001MNRAS.321..398M,2012ApJ...760...99P,2021MNRAS.500.1921G,2022MNRAS.512.4993G,2022ApJS..259...25H,2022ApJ...925..178H,2022A&A...661A..61T}.
The secular gravitational effects between the outer companion and inner binary can drive the inner binary to high eccentricity in Zeipel-Lidov-Kozai oscillations \citep{1910AN....183..345V,1962AJ.....67..591K,1962P&SS....9..719L,2016ARA&A..54..441N,2019MEEP....7....1I}, which can enhance mergers of compact objects \citep[e.g.][]{2002ApJ...578..775B,2011ApJ...741...82T,2013MNRAS.430.2262H,2014MNRAS.439.1079A,2017ApJ...841...77A,2017ApJ...846L..11L,2018ApJ...863...68L,2017ApJ...836...39S,2018ApJ...865....2H,2018ApJ...856..140H,2018ApJ...853...93R,2018ApJ...864..134R,2018A&A...610A..22T,2019MNRAS.486.4443F,2022MNRAS.516.1406S}.
For example, dynamical effects of the outer companion can enhance the merger rate of inner compact double white dwarf binaries relative to isolated binary systems \citep{2011ApJ...741...82T}.
Destabilized stellar triple systems in the field can eject a component with runaway speeds of tens of $\mathrm{km}~\mathrm{s}^{-1}$ \citep{2022A&A...661A..61T}.

Unstable triple systems in star clusters can also result in the ejection of the tertiary perturber and escape from the cluster \citep{2001MNRAS.321..398M}.
From their $N$-body simulations of open clusters, \citet{2001MNRAS.321..398M} describe a particular event in which an initially stable triple system became unstable and the tertiary companion was ejected; the resulting binary and single star both escaped the cluster with respective terminal velocities of $20.6$ and $65.7~\mathrm{km}~\mathrm{s}^{-1}$.
Star clusters are known to harbour exotica such as blue stragglers, which can be formed through the dynamical evolution of stellar triple systems \citep{1999ASPC..169..432I} and have been found in both open clusters \citep{2005BAAA...48..177A,2007A&A...463..789A,2006A&A...459..489D} and globular clusters \citep{2012Natur.492..393F,2016ApJ...830..139P}.
Furthermore, cluster members can display different kinematic relations than what is expected for a disc in equilibrium, and many nearby clusters are moving at speeds of tens of $\mathrm{km}~\mathrm{s}^{-1}$ relative to the local standard of rest \citep{2018A&A...616A..10G,2018A&A...619A.155S,2019A&A...623C...2S,2019A&A...628A..66L}.
The kinematically anomalous population we observe could thus conceivably originate from old open clusters or globular clusters that are close to the Galactic disc.
While open clusters are typically younger than the members of the anomalous population, some open clusters are old enough to be a source of these white dwarfs, and those old open clusters are also typically rich \citep{2018A&A...616A..10G,2018A&A...619A.155S,2019A&A...623C...2S}.
As open clusters are only loosely gravitationally bound, stars can also readily escape from them; for example, observations indicate that the Pleiades has lost $\sim 20$ per cent of its mass over the past 100~Myr \citep{pleiades}.
Though globular clusters are more tightly bound than open clusters and thus less prone to the escape of cluster members, they are also typically much more populated \citep{2018A&A...616A..10G} and only a small fraction of escapees is needed to explain the anomalous population that we observe in this work.

In order to explain the clustering in photometric age in this scenario, the fast-moving population must be produced through a singular event rather than some continuous process.
This could be cluster evaporation in the case of an open cluster origin, or it could be some transient event that disrupted a globular cluster. 
This transient event may relate to the burst of star formation in the local Galactic disc that peaked around $2-3~\mathrm{Gyr}$ ago and continued until $\sim 1~\mathrm{Gyr}$ ago \citep{Mor2019}.
A travel time of around $1~\mathrm{Gyr}$ from the location of origin after the disruption of the cluster would then produce the observed clustering in age of the fast-moving population.

\section{Conclusions} \label{sec:conclusions}

As a follow-up to our previous work on the distribution of photometric cooling ages of massive white dwarfs in \gaia\ EDR3 \citepalias{massive}, we analysed the kinematics of the transverse motion of $0.95-1.25~M_\odot$ \gaia\ EDR3 white dwarfs.
We find that the population of anomalously fast ultramassive white dwarfs on the Q branch reported by \citet{Cheng2019} for \gaia\ DR2 is still present in \gaia\ EDR3.
These white dwarfs appear photometrically young according to single stellar evolution models, but are moving faster than expected for stars of that age according to AVRs observed for the local thin disc.
Using white dwarf cooling models for which the photometric age distributions are consistent with the expectation from the star formation history \citepalias{massive} to infer the masses and ages of our sample of white dwarfs and sorting our full sample into mass and age bins, we find that this population of fast-movers is concentrated to masses of $1.15-1.25~M_\odot$ and photometric cooling ages of $0.5-1.5$~Gyr.

Our analysis of the distributions of the individual components of the transverse velocity reveals that among the white dwarfs in the mass range $1.15-1.25~M_\odot$ and age range $0.5-1.5$~Gyr, there is a population of white dwarfs lagging the local Galactic rotation that is too dispersed in $V_\perp$ to be explained solely by a cooling delay in white dwarfs born in the local thin disc. According to the AVRs determined from observations of main-sequence stars in the local thin disc \citep{Holmberg2009}, it would take longer than the age of the thin disc (10 Gyr) for disc heating to produce a dispersion large enough to explain this anomalous population of white dwarfs.
Furthermore, we find this population to be more dispersed in $V_\perp$ than in $U_\perp$, which suggests that this population does not originate from the local disc.
Some potential explanations of this population include a halo origin, in conjunction with an extra cooling delay or from a dynamically distinct population, or that they are produced through triple system dynamics such as the merger of the inner binary of a hierarchical triple system whose tertiary companion was ejected resulting in a large velocity for the binary that ultimately merges to form the white dwarf.
However, the precise origin of this population of young ultramassive white dwarfs is an open question for future work.

\section*{Acknowledgements}

This work has been supported by the Natural Sciences and Engineering Research Council of Canada through the Discovery Grants program and Compute Canada. I.C. is a Sherman Fairchild Fellow at Caltech and thanks the Burke Institute at Caltech for supporting her research.

This work has made use of data from the European Space Agency (ESA) mission \gaia\ (\url{https://www.cosmos.esa.int/gaia}), processed by the \gaia\ Data Processing and Analysis Consortium (DPAC, \url{https://www.cosmos.esa.int/web/gaia/dpac/consortium}). Funding for the DPAC has been provided by national institutions, in particular the institutions participating in the \gaia\ Multilateral Agreement.
We used the ``Synthetic Colors and Evolutionary Sequences of Hydrogen- and Helium-Atmosphere White Dwarfs'' website at \url{http://www.astro.umontreal.ca/~bergeron/CoolingModels/}.


\section*{Data Availability}

The data used in this paper are available through TAP Vizier and the \gaia\ archive.  
We constructed the 200-pc white dwarf catalogue from \url{https://warwick.ac.uk/fac/sci/physics/research/astro/research/catalogues/gaiaedr3_wd_main.fits.gz}.
The corresponding author can also provide software to perform the analysis presented in this paper.



\bibliographystyle{mnras}
\bibliography{main}


\bsp	
\label{lastpage}
\end{document}